\documentclass[english, onecolumn, aps, prd, 10pt]{revtex4-2}
\usepackage[T1]{fontenc}
\usepackage{stackrel}
\usepackage{graphicx} 
\usepackage[utf8]{inputenc}
\usepackage{subcaption}
\usepackage{amsmath}
\usepackage{amssymb}
\usepackage[serbian]{babel}
\usepackage{float}
\usepackage{tgtermes}
\usepackage{setspace}
\usepackage{braket}
\usepackage{siunitx}
\usepackage[hidelinks]{hyperref}
\usepackage{mathtools}
\usepackage{stackrel}

\begin{document}

\title{If Quantum Measurements Are Secretly Continuous Nonunitary Processes, Weak
Measurements Can Detect It}
\author{Igor Prlina}
\email{prlina@ipb.ac.rs}
\affiliation{Institute of Physics, University of Belgrade, Pregrevica 118, 11080 Belgrade, Serbia}

\author{Milutin Živković}
\affiliation{University of Niš, Univerzitetski trg 2, 18104 Niš, Serbia}
\begin{abstract}
The standard approach to quantum measurements is to assume that they lead to effectively instantaneous collapse of the quantum state. However, if we assume that we are unable to enforce  at what exact moment of time the measurement occurs due to a finite resolution of any time measurement device, at the level of the ensemble, the measurement would lead to an effectively nonunitary evolution involving a mixed state. Each individual ensemble member would face an instantaneous collapse at different moments of time. This process is completely indistinguishable from fundamental nonunitary evolution at the level of each individual ensemble member, within the framework of strong projective measurements. In this paper, we show that weak postselected measurements can distinguish these two types of evolution. An experimental protocol for determining the nature of quantum collapse is described, and the example of a hydrogen atom is analyzed in detail. 
\end{abstract}
\maketitle
	
	\section{Introduction}
Ever since the early days of quantum mechanics, the projective collapse of the wavefunction in the act of measurement was considered controversial. Many different alternatives of quantum mechanics were formulated in response. One of these alternatives are the so-called objective collapse theories \citep{PenroseCollapse, GhirardiCollapse, DiosiCollapse, ObjectiveCollapseNew}, which assume that the Schrödinger equation must be modified in such a way that the natural evolution of the system mimics projective collapse under appropriate conditions, usually depending on the size of the quantum systems. These modifications usually involve nonlinearity, nonunitarity and random parameters \citep{NonlinearCollapse}. Many of these theories have been significantly constrained by modern experiments \citep{CollapseTest}.

In addition to strong projective measurements, another type of measurements is at our disposal: weak (postselected) measurements \citep{Spin100, WeakMeasurement}. Weak measurements are experimental protocols where a quantum system is first weakly coupled to a delocalized quantum pointer, and then it undergoes postselection, a strong selective projective measurement. Only the quantum pointers that were coupled to ensemble members that have survived postselection are kept, and the expectation value of the quantum pointers among the survivors gives the so-called weak value.  The primary importance of weak measurements is that it can suppress noise relative to the signal, making it useful in many experimental applications \citep{WeakValueAmplify}. However, the fact that the weak interaction does not lead to a collapse of the wavefunction enables probing into fundamentally important questions as well. Weak measurements have been used to measure the wavefunction directly \citep{WavefunctionMeasure}, as well as to find ``trajectories'' assigned to particles in the double slit experiment \citep{WeakTrajectories}.

It has been shown recently that there exists at least one potential high energy phenomenon that cannot be observed using strong projective measurements, but can be observed via weak measurements that involve time averaging (high frequency correlations hidden in mixed states) \citep{QuickOscillations}. Systems undergoing nonunitary evolution have been studied under the framework of postselected strong measurements \cite{NonunitaryABL}. The formalism for weak measurements of systems that undergo nonunitary evolution also exists, and weak measurements have already applied to many nonunitary systems \citep{Cavity, DecoherenceWeak}. This has motivated us to study whether a potential continuous nonunitary objective collapse can be observed using time averaged weak measurements. As we will show in this work, the answer is yes.

The layout of the paper is as follows.  In Section II we give a brief overview of both projective and weak measurements in quantum mechanics. In Section III we present a simple objective collapse model which is, at the level of strong nonpostselected measurements,  indistinguishable from instantaneous projective collapse with finite resolution of the moment of time at which the strong measurement occurred. In Section IV we develop a protocol under which we attempt to conduct a strong projective measurement and the weak coupling of the system to a delocalized quantum pointer simultaneously. We show that, after time averaging, this protocol can distinguish instantaneous projective collapse with finite time resolution from fundamental continuous nonunitary objective collapse. In Section V we illustrate this protocol on the simple yet important example of the hydrogen atom.  Finally, we give our concluding remarks in Section VI, as well as comments on possible future lines of investigation.

	\section{Measurements in Quantum Mechanics}
	
	\subsection{Strong measurements}
	
	In standard quantum mechanics, von Neumann introduced the so-called projection postulate to describe measurements. Suppose the ensemble is in the mixed state $\hat{\rho}$ just before a measurement of an observable $\hat{A}$. The observable $\hat{A}$ can be represented as:
	
	\begin{align}\label{spectral}
		\hat{A}=\sum a_n\hat{\Pi}_n
	\end{align}
	\noindent
	where $a_n$ are the (possibly degenerate) eigenvalues of $\hat{A}$, $\hat{\Pi}_n$ are their corresponding projectors.
	The projection postulate states: the state of the ensemble after a selective measurement into the eigenvalue $a_i$ is:
	
	\begin{align}
		\hat{\rho}'=\frac{\hat{\Pi}_i\hat{\rho}\hat{\Pi}_i}{\text{Tr}[\hat{\Pi}_i\hat{\rho}\hat{\Pi}_i]}
	\end{align}
	
	If the eigenvalue $a_i$ is non-degenerate, it can be shown that the previous equation reduces to:
	
	\begin{align}
		\hat{\rho}'=\hat{\Pi}_i.
	\end{align}
	We can say that the process of measurement ``projects'' or ``collapses'' the initial state $\hat{\rho}$  into a new state $\hat{\rho}'$; this collapse is instantaneous according to the standard formalism, and cannot be accounted for by any unitary evolution. The postulate does not define what constitutes a measurement. There are certain proposals to modify (or even eliminate) this postulate, but such proposals need to fulfill the following conditions: their predictions must be consistent with the von Neumann statistics for all experiments already carried out, and yet must be falsifiable so that the predictions can in principle differ from the standard picture in some cases. The first condition is relatively easy to satisfy, but the second one is problematic. 
	
	We propose that one way to differentiate between the von Neumann postulate and other non-standard suggestions is the use of so-called weak measurements and weak values which undergo time averaging. We will first give an overview of the theory of weak values before introducing a simple toy model describing a continuous nonunitary objective collapse. It will be shown that such a model leads to a different prediction of a time-averaged weak value than the standard von Neumann collapse. We do not claim that this toy model corresponds to the nature of reality, but simply show that weak measurements can in principle verify or falsify at least some objective collapse models.

	\subsection{Weak measurements and weak values}

	Assume we have a quantum ensemble, with which we associate a Hilbert space $H$. First, at time $t_\text{in}$ we preselect (prepare) the given ensemble into the mixed state $\hat{\rho}_\text{in}$, then at a later time $t_w$ we weakly couple some observable $\hat{O}$ of the system to a quantum pointer, and finally at time $t_\text{fin}$ we postselect into the mixed state $\hat{\rho}_\text{fin}$. The weak value of the observable $\hat{O}$, $O_w$, corresponds to the shift in position of a quantum pointer and can be expressed as:
	
	\begin{align}\label{weak}
		O_w=\frac{\text{Tr}\left[\hat{\rho}_2(t_w)\hat{O}\hat{\rho}_1(t_w)\right]}{\text{Tr}\left[\hat{\rho}_2(t_w)\hat{\rho}_1(t_w)\right]}.
	\end{align}
	Here $\hat{\rho}_1(t)$ is the state of the ensemble at time $t$, while $\hat{\rho}_2(t)$ is introduced as an auxiliary backwards evolution of the postselected state, and does not necessarily have direct physical significance. If the preselection and postselection conditions are chosen to correspond to pure states, so that $\hat{\rho}_1(t_w)=\ket{\psi_1(t_w)}\bra{\psi_1(t_w)}$ and $\hat{\rho}_2(t_w)=\ket{\psi_2(t_w)}\bra{\psi_2(t_w)}$, the above formula reduces to the more well-known expression
	
	\begin{align}
		O_w=\frac{\braket{\psi_2(t_w)|\hat{O}|\psi_1(t_w)}}{\braket{\psi_2(t_w)|\psi_1(t_w)}}.
	\end{align}
	Since $\hat{\rho}_1(t)$ is the physical evolution of the ensemble, we can always find its value at time $t_w$. We will use superoperator formalism to determine $\hat{\rho}_2(t_w)$.
	
	The set of all linear operators of the form: $\hat{T}:H\to H$ forms a vector space $L(H)$. We can then define a superoperator as a linear operator from the space $L(H)$ to itself, $\hat{\hat{K}}: L(H)\to L(H)$, where linearity is defined as
	
	\begin{align}\label{linear}
		\hat{\hat{K}}\left(\alpha_1 \hat{A}+\alpha_2\hat{B}\right)=\alpha_1 \hat{\hat{K}}\hat{A}+\alpha_2\hat{\hat{K}}\hat{B},
	\end{align}
	for arbitrary complex numbers $\alpha_1,\alpha_2$, and for any linear operators $\hat{A}, \hat{B}$.	
	An inner product on the space $L(H)$ can be defined as:
	
	\begin{align}\label{inner}
		\braket{\hat{A}|\hat{B}}=\text{Tr}[\hat{A}^\dagger \hat{B}].
	\end{align}
	Analogously to the case of adjoint operators, we may introduce the adjoint superoperators as
	
	\begin{align}\label{adjoint}
		\braket{\hat{A}|\hat{\hat{K}}\hat{B}}=\braket{\hat{\hat{K}}^\dagger \hat{A}|\hat{B}}.
	\end{align}
	We define an evolution superoperator between any two moments of time $t_1$ and $t_2$, $\hat{\hat{E}}(t_1,t_2)$, as follows
	
	\begin{align}\label{definicija}
		\hat{\hat{E}}(t_1,t_2)\hat{\rho}_1(t_1)=\hat{\rho}_1(t_2).
	\end{align}
	The state $\hat{\rho}_2(t_w)$ can be defined using the auxiliary retrograde evolution superoperator $\hat{\hat{R}}(t_w,t_\text{fin})$,
	
	\begin{align}
		\hat{\rho}_2(t_w)=\hat{\hat{R}}(t_w,t_\text{fin})\hat{\rho}_\text{fin}.
	\end{align}
	In principle, the auxiliary retrograde evoltion superoperator does not need to be invertible. Equation \eqref{weak} holds under the assumption that the following condition is met \citep{Cavity},
	
	\begin{align}\label{auxil}
		\hat{\hat{R}}(t_w,t_\text{fin})=\hat{\hat{E}}^{\dagger}(t_w,t_\text{fin}).
	\end{align}

	Let us quickly sketch how this condition is derived for nonunitary evolution. If postselection occurs immediately after the moment of weak interaction, the expression \eqref{weak} trivially holds for both unitary and nonunitary evolution, given that no evolution took place at all. If the evolution is unitary, it is simple to describe the case when postselection occurs some time after the the moment of weak interaction. Namely, given that unitary evolution preserves traces, instead of propagating operators forward in time, one can instead propagate the postselected state backwards in time under the inverse of the same unitary evolution. The situation becomes more complex when the evolution is nonunitary \citep{Cavity}. The postselection occurs at time $t_\text{fin}$ and the state of the pointer needs to be evolved and projected at that point in time into the state $\hat{\rho}_{\text{fin}}$. As such, the expression for the weak value becomes

	\begin{align}\label{weakEvolved}
		O_w=\frac{\text{Tr}\left[\hat{\rho}_{\text{fin}}\hat{\hat{E}}(t_w,t_\text{fin})(\hat{O}\hat{\rho}_1(t_w))\right]}{\text{Tr}\left[\hat{\rho}_{\text{fin}}\hat{\hat{E}}(t_w,t_\text{fin})(\hat{\rho}_1(t_w))\right]}.
	\end{align}
	Using the definition of the adjoint \eqref{adjoint}, the expression becomes

	\begin{align}\label{weakEvolvedAdjoint}
		O_w=\frac{\text{Tr}\left[\hat{\hat{E}}^{\dagger}(t_w,t_\text{fin})(\hat{\rho}_{\text{fin}})\hat{O}\hat{\rho}_1(t_w)\right]}{\text{Tr}\left[\hat{\hat{E}}^{\dagger}(t_w,t_\text{fin})(\hat{\rho}_{\text{fin}})\hat{\rho}_1(t_w)\right]}.
	\end{align}
	We see that under the condition \eqref{auxil} and the definition \eqref{definicija}, the expression \eqref{weakEvolvedAdjoint} reduces to the expression \eqref{weak}.

	Therefore, in order to determine $\hat{\rho}_2(t_w)$, we first need to calculate $\hat{\hat{E}}(t_w,t_\text{fin})$. The equation \eqref{definicija} gives us
	
	\begin{align}\label{nzm}
		\hat{\hat{E}}(t_w,t_{\text{fin}})\hat{\rho}_1(t_w)=\hat{\rho}_1(t_{\text{fin}}).
	\end{align}		
	Note that this equation, as well as the knowledge of $\hat{\rho}_1(t_w)$ and $\hat{\rho}_1(t_{\text{fin}}) $ are not enough to calculate $\hat{\hat{E}}(t_w,t_{\text{fin}})$. Namely, in order to completely determine a  superoperator, we need to know how it acts on all elements of $L(H)$. 
	In order to do that, we first turn our attention to the following condition,
	
	\begin{align}\label{druga}
		\hat{\hat{E}}(t_w,t_\text{fin})\hat{\hat{E}}(t_\text{in},t_w)=\hat{\hat{E}}(t_\text{in},t_\text{fin}).
	\end{align}
	Thus, if we are able to calculate $\hat{\hat{E}}(t_\text{in},t_w)$ and $\hat{\hat{E}}(t_\text{in},t_\text{fin})$, we can determine $\hat{\hat{E}}(t_w,t_\text{fin})$. 

The definition of these superoperators contains the state $\hat{\rho}_\text{in}$,
	
	\begin{align}\label{def1}
		\hat{\hat{E}}(t_\text{in},t_w)\hat{\rho}_\text{in}=\hat{\rho}_1(t_w),
	\end{align}	
	
	\begin{align}\label{def2}
		\hat{\hat{E}}(t_\text{in},t_\text{fin})\hat{\rho}_\text{in}=\hat{\rho}_1(t_\text{fin}).
	\end{align}
	Since this state corresponds to our choice of preselection, we can freely change it and see the action of those superoperators on any appropriate state operator. However, not all operators are possible quantum states; rather, only non-negative (positive-semidefinite) operators of trace one can be realized as quantum states (operators which satisfy these conditions are called density operators). Nevertheless, we do not need to actually check the action of a superoperator on any operator. Since superoperators obey condition \eqref{linear}, then it is enough to determine how the superoperator acts on the elements of a basis of $L(H)$. It can be shown that it is always possible to choose a basis of $L(H)$, such that all basis elements are density operators. In the finite-dimensional case the proof is as follows. Every operator can be written as a linear combination of two Hermitian operators. Since any Hermitian operator can be decomposed as the difference of two non-negative operators, we can see that the set of all non-negative operators spans $L(H)$. In a finite-dimensional space, like $L(H)$, every spanning set contains the basis of that space. After constructing such a basis, we can simply normalize each element by dividing it by its trace; note that this is possible because the only non-negative operator with trace zero is the zero operator, which cannot be in the basis. In the infinite-dimensional case, only trace-class operators can be written as a sum of density matrices, but those are the only operators that are relevant to us. Thus, we can determine the superoperators $\hat{\hat{E}}(t_\text{in},t_w)$ and $\hat{\hat{E}}(t_\text{in},t_\text{fin})$ from equations \eqref{def1} and \eqref{def2} by observing how the system evolves from different preselection conditions, which correspond to basis elements of $L(H)$. Finally, we can determine the required superoperator $\hat{\hat{E}}(t_w,t_\text{fin})$ from the condition \eqref{druga}.

	\section{Objective collapse model}
	
	From this point on, we will assume that any systems we consider have negligible evolution, besides the strong measurement process, in the time interval between preselection and postselection.
	In order to show that weak values can in principle lead to falsifiable predictions of objective collapse models, we consider the following simple toy model. Before a strong selective measurement, which begins at time $t'$, of an observable $\hat{A}$ into the non-degenerate eigenvalue $a_i$, the ensemble is in the state $\hat{\rho}(t')$. Such a measurement has the characteristic duration $\Delta t_c$, so that during a measurement over the ensemble, each system separately undergoes a continual collapse according to the rule:
	
	\begin{align}\label{objektiv}
		t\in (t',t'+\Delta t_c)\implies \hat{\rho}(t)=\left(1-\frac{t-t'}{\Delta t_c}\right)\hat{\rho}(t')+\frac{t-t'}{\Delta t_c}\hat{\Pi}_i
	\end{align}
	If characteristic duration $\Delta t_c$ is always much shorter than the current technological limit of measurements of time, it would be impossible to falsify this model. Instead we consider the possibility that the instrumental limit is at most a few orders of magnitude larger than the characteristic duration, and even the case when the true instrumental limit is actually shorter of the two. This model cannot be falsified by strong measurements, which we will now show.
	
	When we use the standard von Neumann approach and claim we perform a strong measurement at time $t'_s$, we actually need to have a large number of systems and equal number of measuring devices, which all turn on exactly at $t'_s$. However, that is not technically achievable due to the finite resolution of our time measuring device (e.g. a stopwatch). If our stopwatch has a resolution of $\Delta t_m$, then we can only be certain that each measurement device is turned on at some point inside the interval $(t'_s-\frac{\Delta t_m}{2},t'_s+\frac{\Delta t_m}{2})$. We can define $t_s=t'_s-\frac{\Delta t_m}{2}$, so the interval becomes $(t_s,t_s+\Delta t_m)$. Then at any moment $t\in (t_s,t_s+\Delta t_m)$ we have one subensemble in which all measuring devices have turned on, and the other subensemble in which no measuring devices have turned on. In the standard von Neumann collapse, the first subensemble would then be described by the state $\hat{\Pi}_i$, and the other by the initial state $\hat{\rho}(t_s)$. The entire ensemble is then in the state:
	
	\begin{align}\label{vn12}
		t\in (t_s,t_s+\Delta t_m)\implies  \hat{\rho}(t)=\tilde{P}(t)\hat{\Pi}_i+ \left(1-\tilde{P}(t) \right)\hat{\rho}(t_s)
	\end{align}
	\noindent
	where $\tilde{P}(t)$ is the probability that a measuring device turns on between $t_s$ and $t$. We have
	
	\begin{align}\label{probab}
		\tilde{P}(t)=\int\limits_{t_s}^{t} P(t')\text{d}t'
	\end{align}
	\noindent
	where $P(t')$ is the probability density, namely the probability that a measuring device turns on at an infinitesimal interval around $t'$. We assume that the probability of a device turning on around any moment of time inside the interval $(t_s,t_s+\Delta t_m)$ is the same, so that $P(t')$ is some constant, $P(t')\equiv P$. Of course, the probability that the device turns on within an interval equal to the resolution of the stopwatch is 1, thus
	
	\begin{align}
		\int\limits_{t_s}^{t_s+\Delta t_m} P\text{d}t=1 \implies P=\frac{1}{\Delta t_m}.
	\end{align}
	Returning to equation \eqref{probab}, we have:
	
	\begin{align}
		\tilde{P}(t)=\frac{t-t_s}{\Delta t_m}.
	\end{align}
	Finally, equation \eqref{vn12} becomes
	
	\begin{align}
		t\in (t_s,t_s+\Delta t_m)\implies  \hat{\rho}(t)=\left(1-\frac{t-t_s}{\Delta t_m}\right)\hat{\rho}(t_s)+\frac{t-t_s}{\Delta t_m}\hat{\Pi}_i.
	\end{align}
	Notice that the projection postulate gives effectively the same ''evolution'' of the ensemble as the continuous nonunitary model \eqref{objektiv}, so that if $\Delta t_m=\Delta t_c$, we will have the same predictions for any strong measurement. We will now show that when time averaged weak measurements are utilized, it is possible to differentiate between the two models, even for $\Delta t_m=\Delta t_c$.

	\section{Simultaneous strong and weak measurements}
	
	In order to propose a way to differentiate the continuous nonunitary model and the instantaneous collapse model, we introduce the following experimental protocol. At time $t_\text{in}$ we preselect the ensemble into the state $\hat{\rho}_\text{in}$. At time $t_m$ we intend to turn on all devices for strong measurements of $\hat{A}$ simultaneously with all the devices for a weak measurement of $\hat{O}$. Due to the finite resolution of our stopwatch $\Delta t_m$, we will fail to achieve simultaneity. A given strong measurement device turns on at some time $t_s\in (t_m-\frac{\Delta t_m}{2},t_m+\frac{\Delta t_m}{2})$, while a weak measurement device turns on at time $t_w$ from the same interval. Finally, at time $t_\text{fin}$, we postselect the ensemble into the state $\hat{\rho}_{\text{fin}}$.
In the next two subsections we will demonstrate that this experimental protocol leads to different time averaged weak values $O_w$ for the two considered models.
	
	\subsection{Weak value with the projection postulate}
	
	The von Neumann postulate requires that individual systems collapse into the same pure state. Due to the finite time resolution, the collapse occurs at different moments of time for different members of the ensemble.  Define $t_m-\frac{\Delta t_m}{2}=0$ for the entire ensemble, so that each strong measurement device turns on at $t_s\in(0,\Delta t_m)$, while each weak measuring device turns on at $t_w\in(0,\Delta t_m)$. We will therefore have a mixture of two subensembles, one in which we have $t_w>t_s$, and the other in which the reverse is true, $t_w<t_s$.
	
	First we consider the subensemble with $t_w>t_s$, i.e. the strong measurement occurs before the weak measurement. Since the strong measurement erases all previous history, any information from the actual preselection condition $\hat{\rho}_{\text{in}}$ is lost. Therefore, the strong measurement of $\hat{A}$ acts like preselection. Strong selective measurement of $\hat{A}$ into the non-degenerate eigenvalue $a_i$ with the corresponding projector $\hat{\Pi}_i$ changes the state of the subensemble to $\hat{\Pi}_i$ at time $t_s$. Since we assumed no evolution outside of strong measurement, equation \eqref{weak} then gives us
	
	\begin{align}\label{exp}
		O_w(t_w>t_s)=\frac{\text{Tr}[\hat{\rho}_\text{fin}\hat{O}\hat{\Pi}_i]}{\text{Tr}[\hat{\rho}_\text{fin}\hat{\Pi}_i]}.
	\end{align}
	
	We now turn our attention to the subensemble where $t_w<t_s$, so that weak measurement happens before the collapse. Now the strong measurement of $\hat{A}$ acts as postselection, while the actual postselection $\hat{\rho}_{\text{fin}}$ only has the effect of lowering the total number of elements of the ensemble that are considered for the statistics. We therefore have
	
	\begin{align}\label{exp2}
		O_w(t_w<t_s)=\frac{\text{Tr}[\hat{\Pi}_i\hat{O}\hat{\rho}_\text{in}]}{\text{Tr}[\hat{\Pi}_i\hat{\rho}_\text{in}]}.
	\end{align}
Now we can average the weak value over the entire ensemble,
	
	\begin{align}
		O_w=\int\limits_{0}^{\Delta t_m}\int\limits_{0}^{\Delta t_m}P_s(t_s)P_w(t_w)O_w(t_s,t_w)\text{d}t_w \text{d}t_s ,
	\end{align}
	\noindent
	where $P_s(t_s)$ is the probability of a strong measuring device turning on at an infinitesimal interval around $t_s$, while $P_w(t_w)$ is the probability of a weak measuring device turning on at an infinitesimal interval around $t_w$. Assuming that these probabilities are constants within the required interval, by a similar argument as before we get

	\begin{align}
		O_w=\frac{1}{\Delta t_m ^2}\int\limits_{0}^{\Delta t_m}\int\limits_{0}^{\Delta t_m}O_w(t_s,t_w)\text{d}t_w \text{d}t_s,
	\end{align}
	
	\begin{align}
		O_w=\frac{1}{\Delta t_m ^2}\left(\int\limits_{0}^{\Delta t_m}\int\limits_{0}^{t_s}O_w(t_w<t_s)\text{d}t_w \text{d}t_s+\int\limits_{0}^{\Delta t_m}\int\limits_{t_s}^{\Delta t_m}O_w(t_w>t_s)\text{d}t_w \text{d}t_s\right).
	\end{align}
	
	By combining the last equation with equations \eqref{exp} and \eqref{exp2}, we obtain
	
	\begin{align}\label{vnw}
		O_w=\frac{1}{2}\left(\frac{\text{Tr}[\hat{\Pi}_i\hat{O}\hat{\rho}_\text{in}]}{\text{Tr}[\hat{\Pi}_i\hat{\rho}_\text{in}]}+\frac{\text{Tr}[\hat{\rho}_\text{fin}\hat{O}\hat{\Pi}_i]}{\text{Tr}[\hat{\rho}_\text{fin}\hat{\Pi}_i]}\right).
	\end{align}

	\subsection{Weak value in the case of the objective collapse model}
	
	The major difference compared to the von Neumann picture is that the nonunitary continuous objective collapse model is not effective, but applies to individual quantum systems directly. We will still consider the weak coupling of the observable $\hat{O}$ to the system to be an instantaneous process, which happens at some point $t_w\in (t_m-\frac{\Delta t_m}{2},t_m+\frac{\Delta t_m}{2})$. However, the objective collapse model describes the strong measurement of $\hat{A}$ as a continuous process occuring for the duration of the interval $(t_m-\frac{\Delta t_c}{2},t_m+\frac{\Delta t_c}{2})$. 
	For each system we will define $t_m-\frac{\Delta t_c}{2}=0$, so the collapse interval becomes $(0,\Delta t_c)$. The equation \eqref{objektiv} gives us
	
	\begin{align}\label{obj}
		t\in (0,\Delta t_c)\implies \hat{\rho}(t)=\left(1-\frac{t}{\Delta t_c}\right)\hat{\rho}_\text{in}+\frac{t}{\Delta t_c}\hat{\Pi}_i
	\end{align}
	According to our redefinition, the weak measurement happens at an instant $t_w\in\left(\frac{1}{2}(\Delta t_c-\Delta t_m),\frac{1}{2}(\Delta t_c+\Delta t_m)\right)$.
	First, we will assume $\Delta t_c\geq \Delta t_m$. Thus, the weak interaction must happen during the collapse interval, and from the previous equation we have
	
	\begin{align}\label{obj2}
		\hat{\rho}_1(t_w)=\left(1-\frac{t_w}{\Delta t_c}\right)\hat{\rho}_\text{in}+\frac{t_w}{\Delta t_c}\hat{\Pi}_i
	\end{align}
	Since we have assumed no evolution besides the collapse process, the condition \eqref{druga} now becomes
	
	\begin{align}\label{centr}
		\hat{\hat{E}}(t_w,\Delta t_c)\hat{\hat{E}}(0,t_w)=\hat{\hat{E}}(0,\Delta t_c).
	\end{align}
	First we consider the following superoperator definition,
	
	\begin{align}
		\hat{\hat{E}}(0,\Delta t_c)\hat{\rho}_\text{in} = \hat{\Pi}_i.
	\end{align}
	From the discussion in the section about weak values in general we recall that we can determine $\hat{\hat{E}}(0,\Delta t_c)$ by choosing preselection conditions equal to elements of a given basis of $L(H)$. Expanding any operator $\hat{\xi}$ in that basis, and using linearity of the trace, we get:
	
	\begin{align}
		\hat{\hat{E}}(0,\Delta t_c)\hat{\xi}=\text{Tr}[\hat{\xi}]\hat{\Pi}_i
	\end{align}
	\noindent
	We will call this kind of map the collapse superoperator $\hat{\hat{C}}$,
	
	\begin{align}\label{trag}
		\hat{\hat{C}}\hat{\xi}=\text{Tr}[\hat{\xi}]\hat{\Pi}_i,
	\end{align}
	\noindent
	and as such,
	
	\begin{align}\label{pretp}
		\hat{\hat{E}}(0,\Delta t_c)=\hat{\hat{C}}
	\end{align}
	Now we can similarly determine $\hat{\hat{E}}(0,t_w)$. We have
	
	\begin{align}
		\hat{\hat{E}}(0,t_w)\hat{\rho}_\text{in}=\left(1-\frac{t_w}{\Delta t_c}\right)\hat{\rho}_\text{in}+\frac{t_w}{\Delta t_c}\hat{\Pi}_i.
	\end{align}
	We can again vary the preselection condition so that it goes through all elements of the chosen basis. In practically identical manner we arrive at
	
	\begin{align}\label{posl}
		\hat{\hat{E}}(0,t_w)=\left(1-\frac{t_w}{\Delta t_c}\right)\hat{\hat{I}}+\frac{t_w}{\Delta t_c}\hat{\hat{C}}.
	\end{align}
	Now, combining equations \eqref{centr}, \eqref{pretp} and \eqref{posl} we have
	\begin{align}\label{supjna}
		\hat{\hat{E}}(t_w,\Delta t_c)\left(\left(1-\frac{t_w}{\Delta t_c}\right)\hat{\hat{I}}+\frac{t_w}{\Delta t_c}\hat{\hat{C}}\right)=\hat{\hat{C}}.
	\end{align}
	
	We now need to solve this superoperator equation for $\hat{\hat{E}}(t_w,\Delta t_c)$. First, notice that from the definition of the superoperator $\hat{\hat{C}}$, equation \eqref{trag}, it immediately follows that $\hat{\hat{C}}^2=\hat{\hat{C}}$. We can then conclude that one possible solution of the previous equation is $\hat{\hat{E}}(t_w,\Delta t_c)=\hat{\hat{C}}$. We now need to check if there are any other solutions. We can start by writing a possible solution in the form
	
	\begin{align}\label{solution}
		\hat{\hat{E}}(t_w,\Delta t_c)=\hat{\hat{C}}+\hat{\hat{B}}
	\end{align}
	\noindent 
	where $\hat{\hat{B}}$ is some superoperator. We will now prove that the only choice of $\hat{\hat{B}}$ which solves the equation \eqref{supjna} is $\hat{\hat{B}}=\hat{\hat{0}}$.
	We substitute equation \eqref{solution} into \eqref{supjna} to obtain:
	
	\begin{align}
		\left(1-\frac{t_w}{\Delta t_c}\right)\hat{\hat{B}}+ \frac{t_w}{\Delta t_c}\hat{\hat{B}}\hat{\hat{C}}=\hat{\hat{0}}.
	\end{align}
	Acting by both sides of the last equation on any operator $\hat{\xi}\in L(H)$, we get
	
	\begin{align}\label{prethodno2}
	\left(1-\frac{t_w}{\Delta t_c}\right)\hat{\hat{B}}\hat{\xi}+ \frac{t_w}{\Delta t_c}\text{Tr}[\hat{\xi}]\hat{\hat{B}}\hat{\Pi}_i=\hat{0}.
	\end{align}
	Since this holds for any operator, for $\hat{\xi}=\hat{\Pi}_i$ we get
	
	\begin{align}
		\hat{\hat{B}}\hat{\Pi}_i=\hat{0}.
	\end{align}
	Substituting this back into \eqref{prethodno2}, we obtain
	
	\begin{align}
		\hat{\hat{B}}\hat{\xi}=\hat{0},
	\end{align}
	\noindent
	or, in other words, $\hat{\hat{B}}=\hat{\hat{0}}$. Therefore, our original solution was indeed unique,
	
	\begin{align}
		\hat{\hat{E}}(t_w,\Delta t_c)=\hat{\hat{C}}.
	\end{align}

	Now we can find the superoperator we actually want. From  the equation \eqref{auxil}, we have
	
	\begin{align}
		\hat{\hat{R}}(t_w,\Delta t_c)=\hat{\hat{E}}^{\dagger}(t_w,\Delta t_c)= \hat{\hat{C}}^{\dagger}.
	\end{align}
	In order to determine the superoperator $\hat{\hat{C}}^\dagger$, we first rewrite the definition \eqref{adjoint},
	
	\begin{align}\label{adjoint2}
		\braket{\hat{A}|\hat{\hat{C}}\hat{B}}=\braket{\hat{\hat{C}}^\dagger \hat{A}|\hat{B}}.
	\end{align}
	From the definition of inner product, equation \eqref{inner}, it follows that
	
	\begin{align}\label{one}
		\braket{\hat{A}|\hat{\hat{C}}\hat{B}}=\text{Tr}[\hat{A}^\dagger (\hat{\hat{C}}\hat{B})].
	\end{align}
	Using the equation \eqref{trag} and the properties of trace, we get
	
	\begin{align}\label{two}
		\text{Tr}[\hat{A}^\dagger (\hat{\hat{C}}\hat{B})]=\text{Tr}[\hat{B}]\text{Tr}[\hat{A}^\dagger \hat{\Pi}_i]=\braket{\text{Tr}[\hat{\Pi}_i \hat{A} ]\hat{I}|\hat{B}}.
	\end{align}
	By combining the equations \eqref{one} and \eqref{two}, it follows that
	
	\begin{align}
		\braket{\hat{A}|\hat{\hat{C}}\hat{B}}=\braket{\text{Tr}[\hat{\Pi}_i \hat{A}]\hat{I}|\hat{B}}.
	\end{align}
	Given that $\hat{A}$ and $\hat{B}$ are arbitrary, using the equation \eqref{adjoint2} we conclude that
	
	\begin{align}
		\hat{\hat{C}}^\dagger\hat{A}=\text{Tr}[\hat{\Pi}_i \hat{A}]\hat{I}.
	\end{align}
	We then find
	
	\begin{align}\label{2}
		\hat{\rho}_2(t_w)=\hat{\hat{C}}^\dagger\hat{\rho}_\text{fin}=\text{Tr}[\hat{\Pi}_i\hat{\rho}_\text{fin}]\hat{I}.
	\end{align}
	By combining equations \eqref{weak}, \eqref{obj2} and \eqref{2}, the weak value of a system weakly interacting at $t_w$ is
	
	\begin{align}\label{nmg1}
		\Delta t_c\geq \Delta t_m \implies O_w(t_w)=\left(1-\frac{t_w}{\Delta t_c}\right) \text{Tr}[\hat{O}\hat{\rho}_\text{in}] + \frac{t_w}{\Delta t_c}\text{Tr}[\hat{O}\hat{\Pi}_i].
	\end{align}
	We need to time average the weak value within the interval $t_w\in \left(\frac{1}{2}(\Delta t_c-\Delta t_m),\frac{1}{2}(\Delta t_c+\Delta t_m)\right)$,
	
	\begin{align}
		\Delta t_c \geq \Delta t_m \implies O_w=\frac{1}{\Delta t_m}\int\limits_{\frac{1}{2}(\Delta t_c-\Delta t_m)}^{\frac{1}{2}(\Delta t_c+\Delta t_m)}O_w(t_w)\text{d}t_w,
	\end{align}
	
	\begin{align}\label{objw1}
		\Delta t_c \geq \Delta t_m \implies O_w=\frac{1}{2}\left(\text{Tr}[\hat{O}\hat{\rho}_\text{in}]+\text{Tr}[\hat{O}\hat{\Pi}_i]\right).
	\end{align}

	Now we turn to the case $\Delta t_m > \Delta t_c$. The interval $t_w\in \left(\frac{1}{2}(\Delta t_c-\Delta t_m),\frac{1}{2}(\Delta t_c+\Delta t_m)\right)$ now has to be separated into three intervals. The first is $t_w\in \left(\frac{1}{2}(\Delta t_c-\Delta t_m),0\right)\equiv I_1$, for which we have the weak interaction before the objective collapse begins. Since the system does not evolve outside the collapse interval, we have a similar result as in equation \eqref{exp2},
	
	\begin{align}
	\Delta t_m > \Delta t_c\implies 	O_w(t_w\in I_1)=\frac{\text{Tr}[\hat{\Pi}_i\hat{O}\hat{\rho}_\text{in}]}{\text{Tr}[\hat{\Pi}_i\hat{\rho}_\text{in}]}
	\end{align}
	In the interval $t_w\in \left(\Delta t_c,\frac{1}{2}(\Delta t_c+\Delta t_m)\right)\equiv I_3$, we have a weak interaction after the objective collapse ends. Using analogous argument as for the first interval, we obtain
	
	\begin{align}
		\Delta t_m > \Delta t_c \implies O_w(t_w\in I_3)=\frac{\text{Tr}[\hat{\rho}_\text{fin}\hat{O}\hat{\Pi}_i]}{\text{Tr}[\hat{\rho}_\text{fin}\hat{\Pi}_i]}.
	\end{align}
	In the interval $t_w\in (0,\Delta t_c)\equiv I_2$, we have an analogous result to \eqref{nmg1},
	
	\begin{align}
		\Delta t_m > \Delta t_c \implies O_w(t_w\in I_2)=\left(1-\frac{t_w}{\Delta t_c}\right) \text{Tr}[\hat{O}\hat{\rho}_\text{in}] + \frac{t_w}{\Delta t_c}\text{Tr}[\hat{O}\hat{\Pi}_i].
	\end{align}
	By averaging the weak value,
	
	\begin{align}
		O_w=\frac{1}{\Delta t_m}\left(\int\limits_{I_1}O_w(t_w\in I_1)\text{d}t_w+\int\limits_{I_2}O_w(t_w\in I_2)\text{d}t_w+\int\limits_{I_3}O_w(t_w\in I_3)\text{d}t_w\right)
	\end{align}
	\noindent
	we get
	
	\begin{align}\label{objw}
		\Delta t_m > \Delta t_c \implies O_w=\frac{\Delta t_m- \Delta t_c}{2 \Delta t_m}\left(\frac{\text{Tr}[\hat{\Pi}_i\hat{O}\hat{\rho}_\text{in}]}{\text{Tr}[\hat{\Pi}_i\hat{\rho}_\text{in}]}+\frac{\text{Tr}[\hat{\rho}_\text{fin}\hat{O}\hat{\Pi}_i]}{\text{Tr}[\hat{\rho}_\text{fin}\hat{\Pi}_i]}\right)+\frac{\Delta t_c}{2\Delta t_m}\left(\text{Tr}[\hat{O}\hat{\rho}_\text{in}]+\text{Tr}[\hat{O}\hat{\Pi}_i]\right).
	\end{align}

By comparing \eqref{objw1} and \eqref{objw}, we see that \eqref{objw}, corresponding to $\Delta t_c \geq \Delta t_m$, reduces into \eqref{objw1}, corresponding to $\Delta t_m > \Delta t_c$, when we substitute  $\Delta t_c =\Delta t_m$. The time interval $\Delta t_m$ is the true time resolution of the time measuring device. However, if  $\Delta t_c > \Delta t_m$, and we falsly believe that the collapse is instantaneous, if we only utilize strong measurements in our experiments, it would appear to us as if the time resolution of the moment of this ``instantaneous'' measurement is actually  $\Delta t_c$ instead of  $\Delta t_m$. As such, we can introduce the apparent time resolution as the longer interval of the two,
\begin{align}\label{ta}
	\Delta t_a =\mathrm{max} \{ \Delta t_m, \Delta t_c \} .
	\end{align}
Thus, we can unify the equations \eqref{objw1} and \eqref{objw} in the following form,
\begin{align}\label{objwunify}
		 O_w=\frac{\Delta t_a- \Delta t_c}{2 \Delta t_a}\left(\frac{\text{Tr}[\hat{\Pi}_i\hat{O}\hat{\rho}_\text{in}]}{\text{Tr}[\hat{\Pi}_i\hat{\rho}_\text{in}]}+\frac{\text{Tr}[\hat{\rho}_\text{fin}\hat{O}\hat{\Pi}_i]}{\text{Tr}[\hat{\rho}_\text{fin}\hat{\Pi}_i]}\right)+\frac{\Delta t_c}{2\Delta t_a}\left(\text{Tr}[\hat{O}\hat{\rho}_\text{in}]+\text{Tr}[\hat{O}\hat{\Pi}_i]\right).
	\end{align}
 Finally, let us note that if the collapse interval is much shorter than the resolution of the time measuring device,
	
	\begin{align}
		\Delta t_m \gg \Delta t_c \implies O_w \approx \frac{1}{2}\left(\frac{\text{Tr}[\hat{\Pi}_i\hat{O}\hat{\rho}_\text{in}]}{\text{Tr}[\hat{\Pi}_i\hat{\rho}_\text{in}]}+\frac{\text{Tr}[\hat{\rho}_\text{fin}\hat{O}\hat{\Pi}_i]}{\text{Tr}[\hat{\rho}_\text{fin}\hat{\Pi}_i]}\right),
	\end{align}
	\noindent
	the time-averaged weak value is practically indistinguishable from the value obtained from the projection postulate, equation \eqref{vnw}.

By comparing the weak value of the instantaneous projection case \eqref{vnw} with the objective collapse toy model case \eqref{objwunify}, we see that our protocol can indeed determine if the apparent time resolution in the act of measurement is purely instrumental or purely fundamental due to objective collapse. The protocol can also determine if both the instrumental error and fundamental collapse contribute to the apparent time resolution, as long as the objective colapse contribution is not negligible compared to the instrumental error contribution. We will illustrate the application of protocol in the next Section, by applying it to the case of the hydrogen atom. 
	
	\section{The Hydrogen Atom}
	
	In order to demonstrate that the experimental protocol discussed in the previous section is possible in principle, we will describe how a specific realization of the protocol could look like by uzing the hydrogen atom as the example. We start by preparing hydrogen atoms in the pure state 1s. Hyperfine structure of the hydrogen atom consists of the small difference between the energy of a hydrogen atom in which the proton and electron spins are parallel, compared to the case where they are antiparallel. The atoms should be arranged such that the proton has spin-up in the chosen z-direction. We can then attempt to simultaneously do a strong measurement of the projection of the atom spin (which is equivalent to the measurement of the electron spin) by directing a pulse of electromagnetic radiation of the appropriate frequency at the atom, while also performing a weak measurement by ensuring that the atom goes through a weak magnetic field inside the Stern-Gerlach apparatus.
	For the preselection and postselection conditions we make the choice
	
	\begin{align}
		\hat{\rho}_\text{in}=\ket{\psi_\text{in}}\bra{\psi_\text{in}},
	\end{align}
	\begin{align}
		\hat{\rho}_\text{fin}=\ket{\psi_\text{fin}}\bra{\psi_\text{fin}},
	\end{align}
	\begin{align}
		\ket{\psi_\text{in}}=a\ket{+}+\sqrt{1-|a|^2}\ket{-},
	\end{align}
	\begin{align}
		\ket{\psi_\text{fin}}=b\ket{+}+\sqrt{1-|b|^2}\ket{-},
	\end{align}
	\noindent
	where $\ket{+}$ and $\ket{-}$ refer to the up and down states of the electron spin in the z-direction respectively, while $a,b$ are complex scalars with norm not greater than one, which can be changed during preparation.
	For the strong selective measurement we use laser absorption into parallel spins, so that $\hat{\Pi}_i=\ket{+}\bra{+}$. The Stern-Gerlach apparatus contains weak magnetic field in the z-direction, which corresponds to weak measurements of the observable $\hat{O}=\hat{S}_z$. After a short calculation, we get
	\begin{align}
		\text{Tr}[\hat{\Pi}_i\hat{O}\hat{\rho}_\text{in}]=\frac{1}{2}\hbar|a|^2,
	\end{align}
	\begin{align}
		\text{Tr}[\hat{\Pi}_i\hat{\rho}_\text{in}]=|a|^2,
	\end{align}
	\begin{align}
		\text{Tr}[\hat{\rho}_\text{fin}\hat{O}\hat{\Pi}_i]=\frac{1}{2}\hbar|b|^2,
	\end{align}
	\begin{align}
		\text{Tr}[\hat{\rho}_\text{fin}\hat{\Pi}_i]=|b|^2.
	\end{align}
	If we arrange that $a\neq 0,b\neq 0$, we see from equation \eqref{vnw} that in the von Neumann case we have
	
	\begin{align}\label{eksp1}
		O_w^{\text{(vN)}}=\frac{1}{2}\hbar.
	\end{align}
	We can also calculate
	
	\begin{align}
		\text{Tr}[\hat{O}\hat{\rho_\text{in}}]=\frac{1}{2}\hbar (2|a|^2-1),
	\end{align}
	\begin{align}
		\text{Tr}[\hat{O}\hat{\Pi}_i]=\frac{1}{2}\hbar,
	\end{align}
	\noindent
	and substitute these values into the two possibilities for weak value under the assumption of the objective collapse model, equations \eqref{objw1} and \eqref{objw},
	
	\begin{align}\label{eksp2a}
		\Delta t_c \geq \Delta t_m \implies O_w^{\text{(obj)}}=\frac{1}{2}\hbar|a|^2,
	\end{align}
	\begin{align}\label{eksp2b}
		\Delta t_m > \Delta t_c \implies O_w^{\text{(obj)}}=\frac{1}{2}\hbar\left(1-\frac{\Delta t_c}{\Delta t_m}(1-|a|^2)\right).
	\end{align}
	
	For $\Delta t_c \geq \Delta t_m$ , by comparing equations \eqref{eksp1} and \eqref{eksp2a}, we see that the predictions for time-averaged weak values for the two models always differ as long as we arrange that $|a|\neq 1$. On the other hand, if $\Delta t_m > \Delta t_c$, the time-averaged weak value will differ from the von Neumann prediction if $|a|\neq 1$ and $\Delta t_m$ is not much greater than $\Delta t_c$.
	It is important to note that we do not claim that this experimental scheme is the most optimal approach to distinguishing the two models. The scheme only serves as a proof of concept that the model can be falsifiable by using time averaged weak measurements with current technology in at least one realistic physical system.

\section{Concluding remarks}
In this paper, we have studied a simple nonunitary continuous model of objective collapse which is, under the framework of strong measurements, indistinguishable from simple presence of finite time resolution of the moment in which an instantaneous projective measurement occurs. We have constructed a protocol involving weak measurements undergoing time averaging which can still differentiate between the two possibilities. This protocol is based on attempting to perform a projective measurement simultaneously with a weak interaction with a delocalized quantum pointer. We have demonstrated this protocol on a simple yet important example of the hydrogen atom.  While the studied objective collapse model was very simple and should be understood only as a toy example, we have demonstrated that weak measurements can offer new insights into the nature of quantum measurements. A study of more elaborate objective collapse models would be a natural topic for future research.

\section{Acknowledgments}

The authors would like to thank Marko Vojinović and Igor Salom for useful discussions. This work was supported by the Ministry of Science, Technological Development and Innovations (MNTRI) of the Republic of Serbia.


\begin{thebibliography}{10}

\bibitem{PenroseCollapse} Penrose, R. (1996). On gravity's role in quantum state reduction. General relativity and gravitation, 28(5), 581-600.

\bibitem{GhirardiCollapse} Ghirardi, G. C., Pearle, P., \& Rimini, A. (1990). Markov processes in Hilbert space and continuous spontaneous localization of systems of identical particles. Physical Review A, 42(1), 78.

\bibitem{DiosiCollapse} Diosi, L. (1987). A universal master equation for the gravitational violation of quantum mechanics. Physics letters A, 120(8), 377-381.

\bibitem{ObjectiveCollapseNew} Mertens, L., Wesseling, M., \& van Wezel,
J. (2024). Stochastic field dynamics in models of spontaneous unitarity
violation. SciPost Physics Core, 7(1), 012.

\bibitem{NonlinearCollapse} Mertens, L., Wesseling, M., Vercauteren, N., Corrales-Salazar, A., \& van Wezel, J. (2021). Inconsistency of linear dynamics and Born's rule. Physical Review A, 104(5), 052224.

\bibitem{CollapseTest} Vinante, A., Mezzena, R., Falferi, P., Carlesso, M., \& Bassi, A. (2017). Improved Noninterferometric Test of Collapse Models Using Ultracold Cantilevers. Phys Rev Lett, 119, 110401.

\bibitem{Spin100} Aharonov, Y., Albert, D. Z., \& Vaidman, L. (1988). How the result of a measurement of a component of the spin of a spin-1/2 particle can turn out to be 100. Physical review letters, 60(14), 1351.

\bibitem{WeakMeasurement} Aharonov, Y., \& Vaidman, L. (1990). Properties
of a quantum system during the time interval between two measurements.
Physical Review A, 41(1), 11.

\bibitem{WeakValueAmplify} Jordan, A. N., Martinez-Rincon, J., \&
Howell, J. C. (2014). Technical advantages for weak-value amplification:
when less is more. Physical Review X, 4(1), 011031. 

\bibitem{WavefunctionMeasure} Lundeen, J. S., Sutherland, B., Patel,
A., Stewart, C., \& Bamber, C. (2011). Direct measurement of the quantum
wavefunction. Nature, 474(7350), 188-191.

\bibitem{WeakTrajectories} Kocsis, S., Braverman, B., Ravets, S.,
Stevens, M. J., Mirin, R. P., Shalm, L. K., \& Steinberg, A. M. (2011).
Observing the average trajectories of single photons in a two-slit
interferometer. Science, 332(6034), 1170-1173.

\bibitem{QuickOscillations} Prlina, I. (2025). If mixed states are secretly quickly oscillating pure states, weak measurements can detect it. Journal of Physics A: Mathematical and Theoretical, 58(47), 475302.

\bibitem{NonunitaryABL} Prlina, I. P., \& Nedeljkovi\'{c}, N. N.
(2015). Time-symmetrized description of nonunitary time asymmetric
quantum evolution. Journal of Physics A: Mathematical and Theoretical,
49(3), 035301. 

\bibitem{Cavity} Wiseman, H. M. (2002). Weak values, quantum trajectories, and the cavity-QED experiment on wave-particle correlation. Physical Review A, 65(3), 032111.

\bibitem{DecoherenceWeak} Ferraz, L. B., Martin, J.,\& Caudano, Y. (2024). On the relevance of weak measurements in dissipative quantum systems. Quantum Science and Technology, 9(3), 035029.

\end{thebibliography}
\end{document}